\documentclass[10pt]{iopart}
\usepackage{graphicx}
\usepackage[square, sort&compress, numbers]{natbib}
\usepackage{iopams}
\usepackage[colorlinks=true, linkcolor=blue, citecolor=blue, urlcolor=blue]{hyperref}

\begin{document}
	
\title{Hot carrier relaxation of Dirac fermions in bilayer epitaxial graphene}

\author{J. Huang$^1$, J. A. Alexander-Webber$^1$, T. J. B. M. Janssen$^2$, A. Tzalenchuk$^{2,3}$, T. Yager$^4$, S. Lara-Avila$^4$, S. Kubatkin$^4$, R. L. Myers-Ward$^5$, V. D. Wheeler$^5$, D. K. Gaskill$^5$, R. J. Nicholas$^1$}

\address{$^1$ Department of Physics, University of Oxford, Clarendon Laboratory, Parks Road, Oxford OX1 3PU, United Kingdom}

\address{$^2$ National Physical Laboratory, Hampton Road, Teddington TW11 0LW, United Kingdom}

\address{$^3$ Department of Physics, Royal Holloway, University of London, Egham TW20 0EX, United Kingdom}

\address{$^4$ Department of Microtechnology and Nanoscience, Chalmers University of Technology, S-412 96 G\"{o}teborg, Sweden}

\address{$^5$ U.S. Naval Research Laboratory, 4555 Overlook Avenue SW, Washington D.C. 20375, USA}

\ead{r.nicholas@physics.ox.ac.uk}

\vspace{10pt}

\begin{indented}
\item\today
\end{indented}

\begin{abstract}Energy relaxation of hot Dirac fermions in bilayer epitaxial graphene is experimentally investigated by magnetotransport measurements on Shubnikov-de Haas oscillations and weak localization. The hot-electron energy loss rate is found to follow the predicted Bloch-Gr\"uneisen power-law behaviour of $T^4$ at carrier temperatures from 1.4 K up to $\sim$100 K, due to electron-acoustic phonon interactions with a deformation potential coupling constant of 22 eV. A carrier density dependence $n_e^{-1.5}$ in the scaling of the $T^4$ power law is observed in bilayer graphene, in contrast to the $n_e^{-0.5}$ dependence in monolayer graphene, leading to a crossover in the energy loss rate as a function of carrier density between these two systems. The electron-phonon relaxation time in bilayer graphene is also shown to be strongly carrier density dependent, while it remains constant for a wide range of carrier densities in monolayer graphene. Our results and comparisons between the bilayer and monolayer exhibit a more comprehensive picture of hot carrier dynamics in graphene systems.
\end{abstract}

\vspace{2pc}

\ioptwocol

\section{Introduction}

The discovery of graphene \cite{r35}, a truly two-dimensional (2D) system in the carbon materials family, has sparked extensive theoretical and experimental research over the last decade, on the physics of the unique chiral Dirac fermions \cite{r36,r37} as well as its potential to become a key element for a wide range of applications \cite{r38}. A significant focus has been on its carrier transport properties and scattering mechanisms. In particular, hot carrier dynamics in graphene has considerable importance in determining the performance of high frequency and high power electronics, high-speed sensors, thermal management of electronic devices, and quantum Hall metrology for accurate measurements under higher temperature and current conditions \cite{r39}.

In monolayer graphene, very high energy loss rates with electron-phonon relaxation times an order of magnitude slower than that of a conventional two dimensional electron gas (2DEG), such as in GaAs/Ga$_{1-x}$Al$_x$As heterojunctions, have been observed \cite{r04,r05,r03,r41}, making monolayer graphene an even more promising candidate for the above mentioned applications. Despite the exceptional electronic properties, the lack of a bandgap in monolayer graphene limits its potential applications. On the other hand, a small energy gap can be opened and continuously tuned by an external electric field in coupled bilayer graphene \cite{r40}, allowing more control and flexibility for technological purposes. Very recently, energy loss rates for hot carriers in bilayer graphene have been theoretically explored taking into account the interactions of electrons with acoustic and surface polar phonons as well as hot phonon effects \cite{r14,r17,r18}, but to date this has yet to be studied in significant detail experimentally.

In this article, we describe experimental investigations of the energy loss dynamics of hot carriers in bilayer epitaxial graphene obtained by magnetotransport measurements, using two independent techniques: firstly from the damping of Shubnikov-de Haas (SdH) oscillations at high magnetic fields and secondly from the suppression of weak localization (WL) peaks in the low magnetic field regime. Energy loss rates, together with electron-phonon relaxation times are extracted in the carrier temperature range of 1.4 K to around 100 K and the carrier density dependence is determined. We then make comparisons between our data and theoretical predictions \cite{r14,r17,r18}, as well as the energy loss behaviour in monolayer graphene \cite{r04,r05,r16}.

\section{Methodology}

\subsection{Sample preperation}

Bilayer graphene was synthesized on semi-insulating (resistivity $>10^9$ $\Omega\cdot$cm) (0001) 6H-SiC that was misoriented less than 0.1 deg from the $(11\bar{2}0)$ direction.  Prior to graphene synthesis, the substrate was etched during the ramp to growth temperature in 5 standard liters per minute (slm) of Pd-purified $\textrm{H}_2$ at 200 mbar. Graphene synthesis was then performed at $1590\,^{\circ}\mathrm{C}$ for 25 minutes in 10 slm of high purity Ar at 100 mbar \cite{r45}. The film was subsequently characterized by x-ray photoelectron spectroscopy (XPS). Using the attenuation of the C 1s and Si 2p signals from the substrate \cite{r46}, the graphene thickness was determined to be approximately 1.5 monolayers, corresponding to 50\% coverage of bilayer graphene within a 400 $\mu$m spot size.  Subsequent optical transmission measurements over micrometer lateral dimensions confirmed the thickness \cite{r47}.

8-leg Hall bars of various sizes were fabricated using electron beam lithography followed by $\textrm{O}_2$ plasma etching and large-area titanium-gold Ohmic contacting. A non-volatile dual-polymer gating technique \cite{r01} using PMMA/MMA and ZEP520A was applied to tune the carrier density in our bilayer epitaxial graphene by UV illumination or corona discharge \cite{r02} at room temperature. Three devices with electron densities of 1.17, 1.90, 2.83 $\times10^{12}$ cm$^{-2}$, and mobilities of 3080, 2055, 1503 cm$^2$V$^{-1}$s$^{-1}$, respectively, measured at 1.4 K, were used in this study. All the devices measured were selected from the bilayer-rich regions and have small dimensions (20-30 microns wide), in order to minimise the effects of long range inhomogeneities. Magnetotransport measurements, as we will show below, confirmed that the devices used in this study had mostly bilayer graphene.

\subsection{Magnetotransport measurements}

Electrical measurements were carried out using a nitrogen and helium cooled Oxford Instruments 21 T superconducting magnet with a variable temperature insert which can provide steady temperatures in the sample environment from 1.4 K up to 300 K. 4-terminal measurements were made using Keithley 2000 DMMs. DC currents higher than 10 $\mu \textrm{A}$ were supplied by a Keithley SMU and currents from 100 nA to 10 $\mu \textrm{A}$ were supplied by a battery-powered constant current source to reduce noise.

\section{Results}

\subsection{Shubnikov-de Haas oscillations and effective mass of electrons in bilayer graphene}

When external energy is added to a system of charge carriers, they will gain energy and heat up if the power loss to the lattice is less than the power input to the system. After the system undergoes an ultrafast quasithermalization within the electron gas via electron-electron interaction in a timescale of tens of femto-seconds, the distribution function of the hot carriers deviates from the original low-energy equilibrium form and manifests itself into a Fermi-Dirac distribution which can be described by an effective temperature $T_e$. The hot carriers then lose energy to the lattice via electron-phonon interaction on a much longer timescale.

To study the hot carrier dynamics for the case of external energy supplied by an electric field, the carrier temperature, $T_e$, as a function of electrical input power is determined. This relationship can be obtained by recording the damping of the amplitudes of Shubnikov-de Haas oscillations \cite{r03,r04,r05,r06} versus current at a fixed low lattice temperature $T_L$, and then comparing those with the measurements under a fixed small current condition while changing the ambient temperature, since the damping is a result of thermal broadening of the hot carrier distribution in both cases. The damped amplitude $\Delta R_{xx}$ (the change in two-dimensional magneto-resistance $R_{xx}$) can be expressed using the Lifshitz-Kosevich formula \cite{r52},
\begin{eqnarray}
\frac{\Delta R_{xx}}{R_{xx}} \propto \frac{\chi}{\sinh\chi}, \label{eq1}
\end{eqnarray}
where $\chi$ is a function of the effective mass $m^*$, the electron temperature and the magnetic field $B$, for a conventional two-dimensional electron gas,
\begin{eqnarray}
\chi =\frac{2\pi^2k_BT_em^*}{\hbar eB}. \label{eq2}
\end{eqnarray}
Eq. \ref{eq2} is based on the assumption that data is taken from the regime where $\frac{\Delta R_{xx}}{R_{xx}}$ is small so that the electric field across the sample and the two-dimensional density of states can both be regarded as constants.

\begin{figure}
	\includegraphics{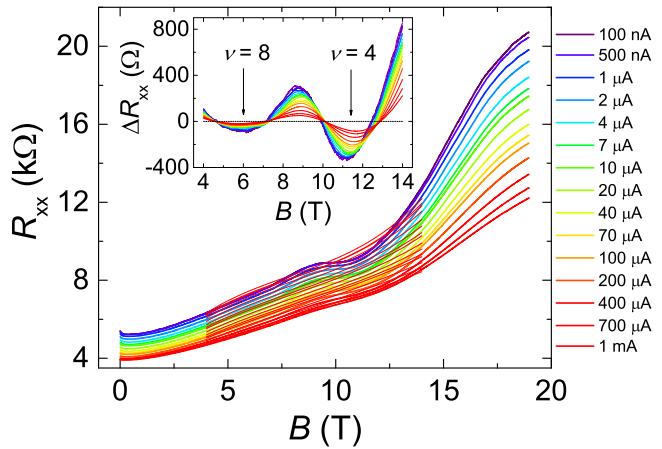}
	\caption{Longitudinal resistance $R_{xx}$ as a function of magnetic field for different input currents ranging from 100 nA to 1 mA, taken at 1.4 K from a bilayer epitaxial graphene sample with carrier density of $1.17\times10^{12}$ cm$^{-2}$. Red lines between 4 T and 14 T are 3rd order polynomials used to subtract the slowly varying background in order to reveal small SdH oscillations. Inset shows the results after the background subtraction.}
	\label{FIG1}
\end{figure}

\begin{figure}
	\includegraphics{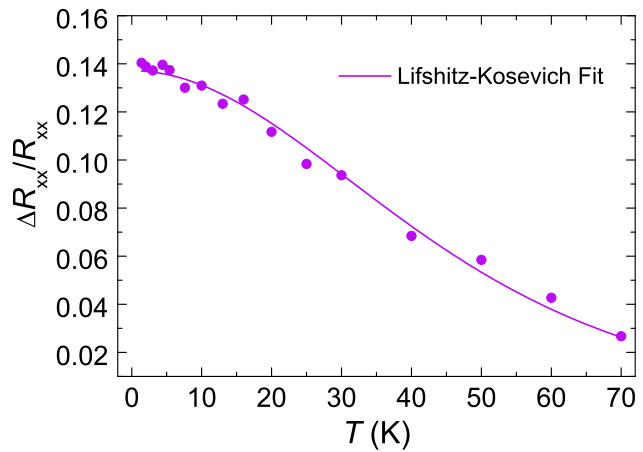}
	\caption{Temperature dependence of the normalized amplitude of SdH oscillations, fitted with Eq. \ref{eq1} and \ref{eq2}, giving a effective mass of 0.033$m_e$ for bilayer epitaxial graphene.}
	\label{FIG2}
\end{figure}

Figure \ref{FIG1} shows the magneto-resistance for a series of currents from 100 nA to 1 mA at a fixed low lattice temperature of 1.4 K, which is effectively anchored by the continuous cooling via helium gas flow in the sample environment. A third order polynomial background subtraction is used to reveal the Shubnikov-de Haas oscillations (Figure \ref{FIG1} inset) from the observed slowly varying background. Two clear resistance minima are observed corresponding to filling factors ($\nu=n_e h / e B$) of 4 and 8, confirming the bilayer nature of the sample \cite{r42}. The amplitudes of the SdH oscillations are found to be strongly damped by increasing input currents. We then measured the damping as a function of ambient temperature, using a constant low current (100 nA) condition to avoid additional carrier heating. Using the Lifshitz-Kosevich formula (Eq. \ref{eq1}), the effective mass $m^*$ of electrons in bilayer graphene is determined to be $0.033m_e$ (for an electron density of $2.83\times10^{12}$ cm$^{-2}$ as an example shown in Figure \ref{FIG2}), which coincides with the value found in previous experiments and theoretical calculations \cite{r07,r08,r09,r10}.

\begin{figure}
	\includegraphics{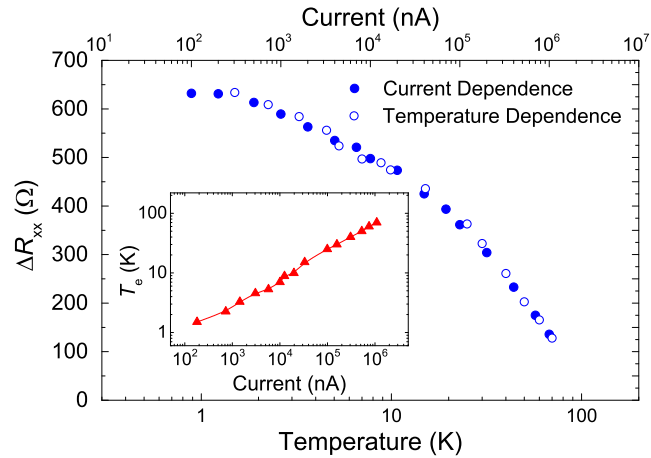}
	\caption{Comparison between the current (full circles) and temperature (open circles) dependences of the amplitudes of SdH oscillations. Inset shows the obtained carrier temperature $T_e$ as a function of injected current.}
	\label{FIG3}
\end{figure}

Comparing the damped amplitudes using both minima and maxima of the SdH oscillations from the current and temperature dependence, the electron temperature as a function of applied current is obtained, shown in Figure \ref{FIG3}. In steady state, the energy loss rate should equal the power input to the system. The energy loss rate per hot carrier for a given electron temperature can be therefore deduced as
\begin{eqnarray}
	P(T_e) = \frac{I^2 R_{xx}}{n_e A}, \label{eq3}
\end{eqnarray}
where $I$ is the applied current, $n_e$ is the carrier density which can be extracted from low field Hall coefficients, and $A$ is the area within the device over which $R_{xx}$ is measured. 

\subsection{Weak localization}

\begin{figure}
	\includegraphics{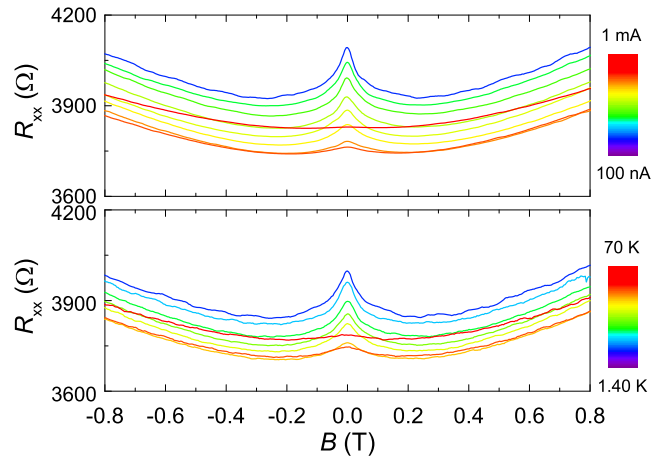}
	\caption{Suppression of the weak localization peaks by increasing current (top) from 100 nA to 1 mA at a fixed $T_L =$ 1.4 K, and by increasing ambient temperature (bottom) from 1.4 K to 70 K at a fixed small current of 100 nA.}
	\label{FIG4}
\end{figure}

Another prominent feature of the magnetotransport is the presence of resistance peaks due to weak localization arising from constructive quantum interference \cite{r43} at low magnetic fields (Figure \ref{FIG4}).
This leads to a second experimental technique to obtain the relationship between the carrier temperature and the corresponding current by measuring the suppression of the weak localization peak heights. This technique has been demonstrated to be extremely helpful in determining the energy loss rates for samples in which SdH oscillations are not observed \cite{r04}. Moreover, it enables a more accessible measurement using magnetic fields less than 1 Tesla.

\begin{figure}
	\includegraphics{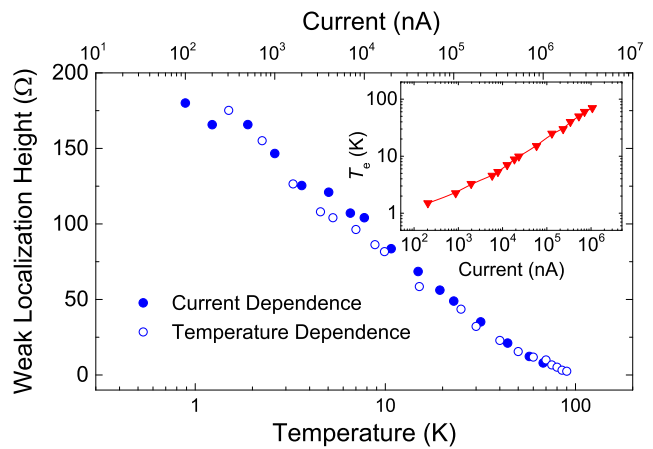}
	\caption{Current (full circles) and temperature (open circles) dependences of the weak localization peak height. Inset shows the $T_e - I$ relationship obtained using this WL method.}
	\label{FIG5}
\end{figure}

The peak height can be calculated as the difference of the longitudinal resistance $R_{xx}$ between 0 T and fixed small magnetic field of 0.2 T. All the current dependence data were taken at a very low lattice temperature of 1.4 K, while all the temperature dependence data were obtained from measurements using a low fixed current of 100 nA. The peak height is significantly suppressed by increasing current or ambient temperature, and the comparison of the suppressed values between these two dependences generates a separate measure of the carrier temperature as a function of applied current (Figure \ref{FIG5}), in addition to the SdH method. 

This method is based on the fact that the amplitude of the weak localization correction to the magneto-resistivity is mainly controlled by the dephasing rate ($\tau_\phi^{-1}$), which is normally considered to be primarily due to electron-electron interactions \cite{r04,r11,r12,r44} at low temperatures and only depends on $T_e$. Therefore, at equilibrium, an increase of the ambient temperature and the electron temperature will have an equal effect on the weak localization. This is also shown in Figure \ref{FIG5}, where the forms of the current and temperature dependences qualitatively agree with each other.

\subsection{Energy loss rate and electron-phonon relaxation time in bilayer graphene}

The energy loss rate as a function of electron temperature is evaluated using Eq. \ref{eq3} from the $I - T_e$ relationships determined by both the SdH and WL techniques. Previous studies including our previous work have found that, in various materials from conventional semiconductors, such as GaAs, to monolayer graphene, the energy loss rates obtained from the two techniques show good correspondence \cite{r04,r11,r13}. Even though the WL is a low-field effect and the SdH oscillations are observed at high magnetic fields, good agreement is also found here for bilayer graphene.

\begin{figure}
	\includegraphics{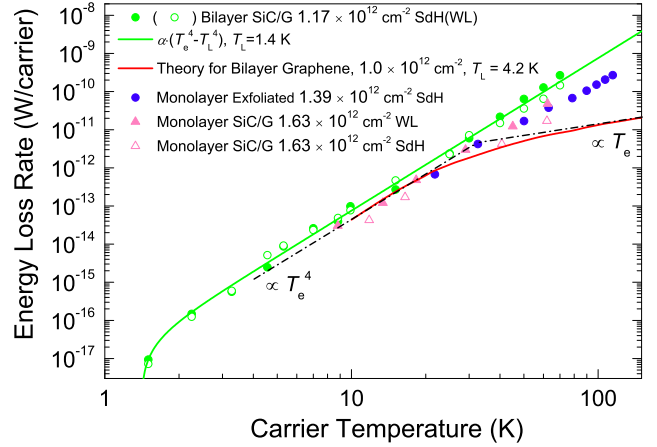}
	\caption{Energy loss rate per carrier as a function of carrier temperature for bilayer epitaxial graphene (green circles) with $n_e = 1.17\times10^{12}$ cm$^{-2}$ measured using both SdH and WL techniques. The energy loss rates are well fitted by the $T^4$ power law shown as the green curve. The red curve is the theoretical prediction from Kubakaddi \cite{r14} for energy loss due to acoustic phonons. The dash-dot lines signify the $T^4$ dependence at low temperatures and the linear-$T$ dependence at very high temperatures. Also shown, for comparison, are the energy loss rates of monolayer graphene with similar carrier densities \cite{r04,r05}.}
	\label{FIG6}
\end{figure}

In Figure \ref{FIG6} the energy loss rates per carrier are shown for the bilayer epitaxial graphene sample with a electron density of $1.17\times10^{12}$ cm$^{-2}$ over a carrier temperature range of 1.4 K to 80 K. The SdH and WL techniques give nearly identical results, which are both well described using a power law dependence
\begin{eqnarray}
	P = \alpha (T_e^4 - T_L^4), \label{eq4}
\end{eqnarray}
where $\alpha$ is a carrier density dependent scaling factor, and the lattice temperature $T_L$ is 1.4 K, resulting in a downward turn of the curve at very low temperatures. The results for bilayer graphene are firstly compared with the power loss in monolayer epitaxial and mechanically exfoliated graphene with similar carrier densities, taken from our previous studies \cite{r04,r05}. For this particular carrier density, hot carriers with a same thermally broadened distribution (same $T_e$) in bilayer graphene lose energy at a rate approximately 2 to 3 times higher than in monolayer graphene. However, as we will show below, the energy loss rate for these two systems have significantly different carrier density dependences.

\begin{figure}
	\includegraphics{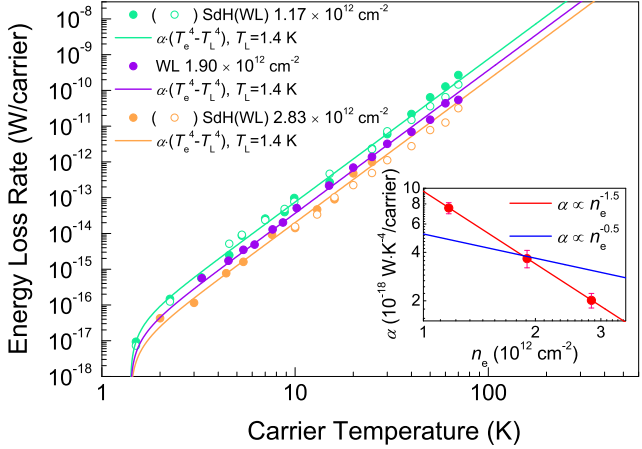}
	\caption{Energy loss rate per carrier versus carrier temperature for bilayer graphene with different carrier densities. Fitted curves are using Eq. \ref{eq5} with $\alpha$ as the fitting parameter. Inset shows the $n_e^{-1.5}$ dependence (red line) of $\alpha$ in bilayer graphene, theoretically predicted by Kubakaddi \cite{r14} with a deformation potential constant of 22 eV. The blue line in the inset is the $n_e^{-0.5}$ dependence \cite{r04,r16} for monolayer graphene. A crossing point at $n_e \approx 1.86\times10^{12}$ cm$^{-2}$ is clearly observed between the bilayer and monolayer dependences.}
	\label{FIG7}
\end{figure}

To further validate the $T^4$ behaviour and study the effects of carrier density on $\alpha$ in bilayer graphene, results from three different carrier densities are compared (Figure \ref{FIG7}) from measurements using both the SdH and WL techniques. The energy loss rates associated with the three different carrier densities all behave in very good agreement with Eq. \ref{eq4}, with different pre-factors. The inset of Figure \ref{FIG7} shows $\alpha$ as a function of carrier density. In contrast with the $T_e^5$ behaviour \cite{r15} of other conventional 2DEG, according to theoretical predictions for the power loss in bilayer graphene by Kubakaddi \cite{r14}, within the low-temperature Bloch-Gr\"{u}neisen (BG) regime, interactions between hot electrons and 2D acoustic phonons indeed give a $T^4$ power law
\begin{eqnarray}
P_{el-ap} = F(T_e) - F(T_L), \label{eq5}
\end{eqnarray}
and
\begin{eqnarray}
F(T) = \frac{{m^*}^2 D^2 (k_BT)^4 3! \zeta(4)}{\pi^{5/2} n_e^{3/2} \rho \hbar^3 (\hbar v_s)^3} = \alpha T^4, \label{eq6}
\end{eqnarray}
where $D$ is the deformation potential constant, $\zeta(4)$ is the Riemann zeta function, $\rho$ is the areal mass density, and $v_s$ is acoustic wave velocity. The theory has an $n_e^{-1.5}$ dependence of $\alpha$, which matches our experimental results very well for the carrier density dependent $\alpha$ using Eq. \ref{eq6} (Figure \ref{FIG7} inset). We emphasize that this $n_e^{-1.5}$ dependence in bilayer graphene, due to its parabolic dispersion relation at low energies \cite{r14}, is very different from the $n_e^{-0.5}$ dependence for monolayer graphene which has been theoretically predicted \cite{r16} and already experimentally observed \cite{r04}. As a comparison, the carrier density dependence of $\alpha$ in monolayer graphene is plotted in the same figure. Below the crossing point corresponding to a carrier density of approximately $1.86\times10^{12}$ cm$^{-2}$ and at carrier temperatures between 1.4 K and 100 K, hot carriers in bilayer graphene will be able to lose energy faster than those in monolayer and vice versa. This therefore explains the power loss difference shown in Figure \ref{FIG6} both qualitatively and quantitatively. It also suggests a higher tunability in the power loss behaviour of bilayer graphene by controlling the carrier density, and could potentially become an important criterion to consider especially for applications where either higher (e.g. ultrafast electronics) or lower (e.g. photo-thermoelectric detectors) energy loss rates are favoured. In addition, the above fittings give a deformation potential constant of 22 eV for our bilayer epitaxial graphene, which is very close to the value used in theoretical studies \cite{r14,r17,r18} and falls in the relatively broad range (10-50 eV) found in the literature \cite{r19,r20,r21,r22}.

It is worth pointing out that the theoretically predicted range in which the $T^4$ power law is valid extends only up to about 20 K before slowing to approach the high-temperature limit of a linear temperature dependence (red curve and dash-dot lines in Figure \ref{FIG6}). Our results, on the other hand, demonstrate a much wider carrier temperature range up to at least 80 K for the $T^4$ dependence. The BG temperatures ($T_{BG}$) corresponding to the three carrier densities for our bilayer graphene are calculated to be 83 K, 114 K and 161 K, suggesting the $T^4$ dependence has not been altered even as the electron temperature approaches $T_{BG}$. This is not commonly seen in conventional 2DEG \cite{r48,r49,r50}, however, very similar behaviour has already been observed in monolayer graphene in a number of experimental studies \cite{r04,r23,r24,r25}. One explanation for this are additional cooling pathways, due to disorder-assisted electron-phonon interaction know as ``supercollisions" \cite{r26} which has been proposed for monolayer graphene, leading to a gradual transition where the energy loss rate changes to a $T^3$ dependence \cite{r23,r24} between the BG regime and the high-temperature regime. So far, no theoretical extension of the ``supercollisions" to bilayer graphene has been reported and our results show no evidence of the transition to a $T^3$ dependence as observed in monolayer graphene. Another possible cooling mechanism to retain the energy loss rate increasing as $T^4$ in a substrate supported bilayer graphene sample could be the interaction between hot electrons and surface polar phonons (SPPs) \cite{r14,r17,r18}, combined with hot phonon effects which occur when the hot phonon decay rate is not as fast as the phonon emission rate. However, theoretical calculations based on a SiC substrate suggest the contribution from SPPs can only be clearly observed for electron temperatures higher than 100 K \cite{r18}, due to the relatively low dielectric constant and high surface polar phonon energies of SiC, compared with substrates such as Hf$\textrm{O}_2$. As a result, more theoretical attention and comparable experimental work on this could prove fruitful. For application purposes, this wide temperature range ($2\sim80$ K) of the $T^4$ dependence may also be advantageous for hot-electron detectors operating at liquid-helium to liquid-nitrogen temperatures.

\begin{figure}
	\includegraphics{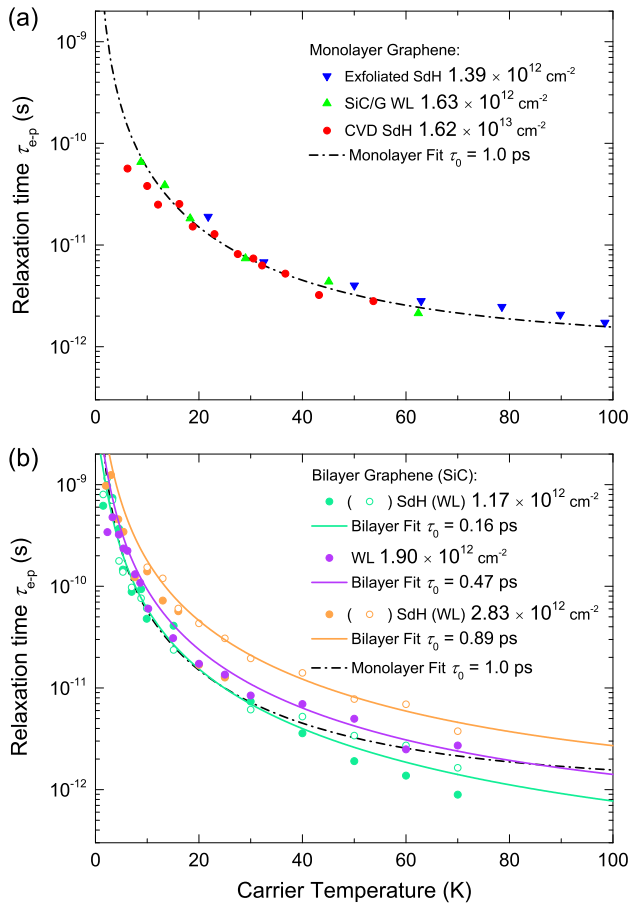}
	\caption{Relaxation time of electron-phonon scattering as a function of carrier temperature between 1.4 K to 100 K. Data shown are deduced from Eq. \ref{eq7} (a) for monolayer graphene with carrier densities spanning over more than an order of magnitude, as well as (b) for bilayer epitaxial graphene with the measured three carrier densities. Fitted lines are using Eq. \ref{eq8} with different phonon relaxation times $\tau_0$.}
	\label{FIG8}
\end{figure}

Another important factor characterizing the hot carrier dynamics is the electron-phonon relaxation time ($\tau_{el-ph}$, also often referred to as the energy loss time). The energy loss time at low temperatures can be generally deduced from the energy loss rate $P$ using the Mott formula through energy balance equations \cite{r16}
\begin{eqnarray}
\tau_{el-ph} = \frac{\pi^2 k_B^2 (p + 1) (T_e^2 - T_L^2)}{6 E_F P}, \label{eq7}
\end{eqnarray}
where $p$ is a constant taking the value of $\sim$1.0 for monolayer and bilayer graphene \cite{r17}. The energy loss time can therefore be plotted as a function of carrier temperature for the three different carrier densities shown in Figure \ref{FIG8}. Also shown for comparison is the determined $\tau_e$ for monolayer graphene of various synthesis methods and a wide range of carrier density from the reported energy loss rates \cite{r04,r05} using Eq. \ref{eq7}.

Through substitution of the energy loss rate given by Eq. \ref{eq4}, the energy loss time is expected to take the form
\begin{eqnarray}
\tau_{el-ph} = \frac{\pi^2 k_B^2 (p + 1)}{6 \alpha E_F (T_e^2 + T_L^2)} + \tau_0. \label{eq8}
\end{eqnarray}
where a limiting phonon relaxation time  $\tau_0$ is added to account for the hot phonon and other high-order effects at high temperatures \cite{r05,r27}. Very good agreement between our experimental results and Eq. \ref{eq8} has been observed as shown in Figure \ref{FIG8} for both the bilayer and previously reported \cite{r04,r05} monolayer graphene. One obvious difference between the bilayer and monolayer cases is that the energy loss time significantly depends on the carrier density in bilayer graphene even though the change in carrier density is only a factor of 2.4. On the contrary, for the monolayer case, the carrier temperature dependence of $\tau_{el-ph}$ seems unchanged for a carrier density variation of over an order of magnitude as shown in Figure \ref{FIG8}(a). This can be partly explained from Eq. \ref{eq8} by noticing that both the scaling factor $\alpha$ and the Fermi level $E_F$ are carrier density dependent. In bilayer graphene, $\alpha \propto n_e^{-1.5}$ \cite{r14} and $E_F \propto n_e$, resulting in $\tau_{el-ph}$ having a net carrier density dependence of $n_e^{0.5}$ at low temperatures. However, in monolayer graphene due to its massless Dirac fermions, $\alpha \propto n_e^{-0.5}$ \cite{r04,r16} and $E_F \propto n_e^{0.5}$ \cite{r28}, leaving $\alpha E_F$ constant over a wide range of $n_e$. 

Another contribution to the observed strong carrier density dependence of $\tau_{el-ph}$ could be from a carrier density dependence of $\tau_0$, as shown in Figure \ref{FIG8}(b). Changing the value of $\tau_0$ affects the shape of the fitting curves more significantly at temperatures above 50 K, consistent with its role in limiting the energy loss rate due to high temperature effects, as discussed above. In addition to its carrier density dependence, $\tau_0$ is predicted to be sensitive to the size of the sample, defects inside the lattice and also edge roughness of the graphene \cite{r30}. Characterizing all the above factors would require sophisticated analysis from measurements on better controlled samples at higher carrier temperatures. Given that the values used in our fits are all within the normal range appearing in the literature \cite{r05,r27,r31,r32,r33,r34}, we would for now only consider $\tau_0$ as a carrier density dependent parameter, which enhances the overall carrier density dependence of $\tau_{el-ph}$.

Thus, the strong carrier density dependence of the electron-phonon relaxation time in bilayer graphene suggests that even faster energy relaxation can be achieved at low carrier densities, which is very important for ultrafast electronics and high-speed communications, as well as for quantum Hall metrology where it plays an important role in limiting the breakdown current \cite{r29}. On the other hand, for some applications, slower electron-phonon relaxation times are preferred for operation. An example of this kind would be detectors based on the photo-thermoelectric effect (PTE). The responsivity of PTE at a given input power directly depends on the resulted carrier temperature \cite{r51}, and thus can be significantly enhanced by longer energy loss time, which, as we have shown, can be easily achieved in bilayer graphene by increasing its carrier density within certain limits. A long energy loss time also suggests a long cooling length \cite{r51}, which would be beneficial to large area PTE devices, or allows higher operating temperatures using bilayer graphene.

\section{Conclusions}

In summary, we have experimentally studied the hot carrier relaxation characteristics in epitaxially grown bilayer graphene by magnetotransport measurements. Both the SdH and WL techniques have been demonstrated to give consistent results. Energy loss rates in bilayer graphene have been found to follow a $T^4$ dependence for carrier temperature from 1.4 K up to $\sim$100 K, and increase with decreasing carrier density as $n_e^{-1.5}$. The electron-phonon relaxation time has also been observed to be carrier density dependent. These behaviours are in good agreement with the theory \cite{r14} accounting for electron scattering due to acoustic phonons at low temperatures. At temperatures above 20 K, the energy loss rates have shown to be much higher than are predicted, with no evidence of approaching the high-temperature linear-$T$ dependence, indicating contributions from other possible cooling mechanisms in this intermediate temperature range, such as the supercollision mechanism \cite{r26}, which has not been explored for bilayer graphene.

Comparisons have been made between bilayer and monolayer graphene. A stronger carrier density dependence ($n_e^{-1.5}$ vs $n_e^{-0.5}$) of the energy loss rate has been confirmed in bilayer graphene, resulting in a crossover point at $n_e \approx 1.86\times10^{12}$ cm$^{-2}$ for energy loss in the two materials. The strong carrier density dependence of $\tau_{el-ph}$ in the bilayer is also in contrast with the carrier density independent behaviour in the monolayer. These relationships can thus provide us with higher tunability of the electron-phonon interactions and the possibility to achieve even faster/slower energy loss for more efficient hot-carrier applications using bilayer graphene.

\ack
This work was supported by the UK EPSRC and NMS, EU Graphene Flagship, EMRP GraphOhm, and also by the US Office of Naval Research.

%\bibliographystyle{iopart-num}
%\nocite{*}
%\bibliography{refs}
\providecommand{\newblock}{}

\end{document}